\documentclass[12pt,reqno]{article}
\usepackage{amssymb,amsthm,amsmath,amsxtra,appendix,bbm}
\usepackage{hyperref} 
\usepackage{mathrsfs}
\usepackage{xcolor}
\usepackage{authblk}

\numberwithin{equation}{section}
\makeatletter

\makeatletter
\def\@tocline#1#2#3#4#5#6#7{\relax
	\ifnum #1>\c@tocdepth 
	\else
	\par \addpenalty\@secpenalty\addvspace{#2}%
	\begingroup \hyphenpenalty\@M
	\@ifempty{#4}{%
		\@tempdima\csname r@tocindent\number#1\endcsname\relax
	}{%
		\@tempdima#4\relax
	}%
	\parindent\z@ \leftskip#3\relax \advance\leftskip\@tempdima\relax
	\rightskip\@pnumwidth plus4em \parfillskip-\@pnumwidth
	#5\leavevmode\hskip-\@tempdima
	\ifcase #1
	\or\or \hskip 1em \or \hskip 2em \else \hskip 3em \fi%
	#6\nobreak\relax
	\dotfill
	\hbox to\@pnumwidth{\@tocpagenum{#7}}
	\par
	\nobreak
	\endgroup
	\fi}
\makeatother

\newtheorem{theorem}{Theorem}[section]
\newtheorem{lemma}[theorem]{Lemma}

\newtheorem{proposition}[theorem]{Proposition}

\theoremstyle{remark}

\renewcommand\phi{\varphi}

\renewcommand{\epsilon}{\varepsilon}

\renewcommand{\ge}{\geqslant}
\renewcommand{\le}{\leqslant}

\renewcommand{\hat}{\widehat}

\newcommand{\bra}[1]{\langle #1|}
\newcommand{\ket}[1]{|#1\rangle}

\usepackage{xargs}
\usepackage[prependcaption]{todonotes}

\newcommandx{\dom}[2][1=]{\todo[inline, author={Dom.}, linecolor=purple,backgroundcolor=purple!15,bordercolor=purple,#1]{#2}}
\newcommandx{\domnote}[2][1=]{\todo[author={Dom.}, linecolor=purple,backgroundcolor=purple!15,bordercolor=purple,#1]{#2}}
\newcommandx{\addom}[2][1=]{\todo[inline, linecolor=purple,backgroundcolor=purple!15,bordercolor=purple,#1]{#2}}

\newcommandx{\nico}[2][1=]{\todo[inline, author={Nicolas}, linecolor=yellow,backgroundcolor=yellow!15,bordercolor=yellow,#1]{#2}}
\newcommandx{\niconote}[2][1=]{\todo[author={Nicolas}, linecolor=yellow,backgroundcolor=yellow!15,bordercolor=yellow,#1]{#2}}
\newcommandx{\addnico}[2][1=]{\todo[inline, linecolor=yellow,backgroundcolor=yellow!15,bordercolor=yellow,#1]{#2}}

\newcommandx{\ale}[2][1=]{\todo[inline, author={Aless.}, linecolor=cyan,backgroundcolor=cyan!15,bordercolor=cyan, #1]{#2}}
\newcommandx{\alnote}[2][1=]{\todo[author={Aless.e}, linecolor=cyan,backgroundcolor=cyan!15,bordercolor=cyan,#1]{#2}}
\newcommandx{\addale}[2][1=]{\todo[inline, linecolor=cyan,backgroundcolor=cyan!15,bordercolor=cyan,#1]{#2}}

\numberwithin{equation}{section}

\begin{document}

\title{Reduced fluctuations for bosons in a double well}

\author{Alessandro Olgiati\thanks{alessandro.olgiati@math.uzh.ch}}
\affil[1]{Institut f\"ur Mathematik, Universit\"at Z\"urich, Winterthurerstrasse 190, 8057 Zurich}

\maketitle

\begin{abstract}
	We review two recent results on the ground state properties of bosonic systems trapped by a double well external potential. In the limit of large number of particles and large separation between the wells we prove that fluctuations in the number of particles occupying each single-well low energy mode occur at a reduced scale with respect to $ \sqrt{N}$, the latter being the typical prediction of the central limit theorem. This signals the breakdown of the independent and uncorrelated particles picture of standard Bose-Einstein condensation and the emergence of an interaction-driven correlated phase in the ground state.
\end{abstract}


\section{Introduction and main results}

The rigorous derivation of effective theories from many-body systems is a well established research line in mathematical physics, which has seen considerable progress in recent years. A typical paradigm, common to many different fields, is the proof of some form of independent particle picture, allowing to reduce a complicated many-body problem to a more tractable one-body one. This occurs both in the classical~\cite{Spohn-12,Golse-13,GalRayTex-14} and quantum~\cite{LieSeiSolYng-05,Schlein-08,BenPorSch-15,Rougerie-EMS} setting. To the latter belongs Bose-Einstein condensation, the physical phenomenon occurring when a Bose gas at extremely low temperature exhibits a macroscopic occupation of a single one-body orbital. The mathematics of the Bose gas has seen a huge amount of important results in the past twenty years: while a rigorous proof of Bose-Einstein condensation for continuous systems in the thermodynamic limit is still lacking, the emergence of the mentioned macroscopic one-body occupation was confirmed in a very diverse range of regimes and settings~\cite{LieSei-02,LieSei-06,NamRouSei-16,BocBreCenSch-19,CarCenSch-21,BreSchSch-21a,DimGia-21}.

Even more recently, considerable progress has been made in the description of fluctuations beyond the macroscopic occupation of the condensate wave-function~\cite{Seiringer-11,GreSei-13,LewNamSerSol-13,DerNap-13,BocBreCenSch-19,NamTri-21,BreSchSch-21b,BasCenOlgPasSch-22}. These effects are accurately captured by Bogoliubov's theory, and are responsible for a number of highly non-trivial next-to-leading order properties (energy expansion, eigenvectors approximation, fluctuations of observables, \dots), the prototypical example of which is the Lee-Huang-Yang second order expansion for the ground state energy~\cite{YauYin-09,FouSol-19,BasCenSch-21,FouSol-21}. For the purpose of our later discussion, it is relevant to mention the series of works~\cite{BenKirSch-13,BucSafSch-13,RadSch-19,Rad-20} which prove that fluctuations of a large class of (expectations of) $N$-body observables have a size of order $\sqrt{N}$ and satisfy a central limit theorem.

Much less common is the study of quantum systems in which correlations influence the leading order behavior by imposing a macroscopic occupation of more than a single one-body orbital. We will address this problem in the present paper, in the particular case of a bosonic $N$-body system trapped by a double-well external potential. The limit of distant wells and high potential barrier, coupled with the $N\to\infty$ limit, forces the emergence of two leading order modes, one for each well. Already in such a simple case, a richer fauna of possible forms of e.g. a good trial state emerges with respect to the fixed (one-)well setting. Indeed, the restriction of a $N$-body model to the span of two one-body modes is still a highly non-trivial $N$-body model (e.g. Hubbard, Bose-Hubbard\dots). This is to be compared with the Hartree model, which captures the leading order behavior of a standard trapped mean-field bosonic system, and is genuinely one-body. It is worth remarking that a macroscopic occupation of multiple one-body modes can be also achieved in multi-species Bose-Einstein condensate~\cite{MicOlg-17a,MicOlg-17b,Olg-17,MicNamOlg-19}. Those systems are however quite far from our setting, which is rather closer to a fragmented condensate~\cite{DimFalOlg-21}. Still, the presence of a small but non-zero gap in the mean-field model (see \eqref{eq:gap}) makes the picture even different from one of exact fragmentation.

The aim of the present paper is the review and discussion of the recent result \cite{OlgRouSpe-21} (and the previous \cite{RouSpe-16}). An amount of information on the associated one-body Hartree model was also proven in the companion paper~\cite{OlgRou-20}. 

In the context of the above two-mode picture for double-well systems, we prove the emergence of important correlations in the ground state by monitoring the behavior of the variance of the number of particles within each well. If $u_1$ is the one-well low energy mode for, say, the right well (having fixed an origin at the midpoint between the two wells) and $\mathcal{N}_{u_1}$ is the associated number operator, then by symmetry arguments the expectation on the ground state should certainly satisfy
\begin{equation*}
\langle \mathcal{N}_{u_1}\rangle_{\psi_\mathrm{gs}} \simeq  \frac{N}{2}\qquad\text{as }N\to\infty
\end{equation*}
irrespectively of whether the two wells really get far apart in the limit. More interesting is the behavior of the variance of such an observable. As already mentioned above, expectations of $N$-body observables on bosonic ground states are typically ruled by a central limit theorem with fluctuations at the scale $\sqrt{N}$~\cite{BenKirSch-13,BucSafSch-13,RadSch-19,Rad-20}, and one should thus expect
\begin{equation*}
\left\langle \left (\mathcal{N}_{u_1}-\frac{N}{2}\right)^2\right\rangle_{\psi_\mathrm{gs}} \simeq O(N)\qquad\text{as }N\to\infty.
\end{equation*}
The justification of this phenomenon is that, if correlations could be neglected, then the operator $\mathcal{N}_{u_1}$ would (essentially) be the sum of $N$ independent and identically distributed random variables, with an expectation of order $N$. As a consequence of CLT, its variance would be of order $N$ as well.

Our main result is that, in a genuine double-well regime (i.e. if the wells get far apart as $N\to\infty$ in such a way that the tunneling is suppressed), the above picture fails, and one instead has
\begin{equation} \label{eq:heuristical_result}
\lim_{N\to\infty}\frac{1}{N}\left\langle \left (\mathcal{N}_{u_1}-\frac{N}{2}\right)^2\right\rangle_{\psi_\mathrm{gs}}=0.
\end{equation}
This is a departure from the prediction of a central limit theorem for fluctuations at the scale $\sqrt{N}$, thus signaling the breakdown of the independent particles picture and therefore the emergence of correlations in the ground state.

Before precisely stating our main result we will introduce all the quantities that will play an important role and we will state the main assumptions.

\subsection{$N$-body model and assumptions}

We consider the $N$-body Hamiltonian
\begin{equation} \label{eq:H_N}
H_N:= \sum_{j=1}^N\big(-\Delta_{x_j}+V_\mathrm{DW}(x_j)\big)+ \frac{\lambda}{{N-1}} \sum_{1\le i<j\le N} w(x_i-x_j)
\end{equation}
on the bosonic Hilbert space
\begin{equation*}
\mathcal{H}_N:=\bigotimes_{\mathrm{sym}}^N L^2(\mathbb{R}^d),
\end{equation*}
with the choice
\begin{equation*}
V_\mathrm{DW}:=\min \big\{ |x-x_L|^s,|x+x_L|^s \big\}, \quad s\ge2
\end{equation*}
and
\begin{equation*}
\pm x_L=\big(\pm L/2,0,\dots,0)\in \mathbb{R}^d.
\end{equation*}

We also assume that $w\in L^\infty(\mathbb{R}^d)$ is radial with compact support, and
\begin{equation*}
w,\hat w \ge0.
\end{equation*}
The parameter $\lambda>0$ in \eqref{eq:H_N} is a coupling constant which we will later suppose to be small enough but fixed.

Since $H_N$ has compact resolvent (due to the fact that $\lim_{|x|\to\infty}V_{\mathrm{DW}}(x)=+\infty$), its spectrum consists of eigenvalues only. In particular, there exists a ground state $\psi_\mathrm{gs}\in \mathcal{H}_N$ such that
\begin{equation*}
H_N \psi_\mathrm{gs} = E(N)\psi_\mathrm{gs},\qquad\text{with}\qquad E(N)=\min \Big\{ \langle \psi_N,H_N \psi_N \rangle,\;|\; \psi_N \in \mathcal{D}[H_N],\;\int_{\mathbb{R}^{dN}}|\psi_N|^2=1 \Big\}.
\end{equation*}
Moreover, $\psi_\mathrm{gs}$ is unique up to an overall phase.

\subsection{Hartree model and low-energy modes}
There is a natural one-body model associated to the $N$-body linear operator $H_N$, obtained by computing expectations of $H_N/N$ on factorized states. This is the well-known Hartree functional
\begin{equation}
\begin{split}
\mathcal{E}^\mathrm{H}[u]:=\;& \int_{\mathbb{R}^d} |\nabla u(x)|^2dx+ \int_{\mathbb{R}^d} V_\mathrm{DW}(x)|u(x)|^2dx\\
&+\frac{\lambda}{2}\iint |u(x)|^2 w(x-y)|u(y)|^2dxdy,
\end{split}
\end{equation}
for $u$ in (a suitable dense subset of) $L^2(\mathbb{R}^d)$. Indeed, $\mathcal{E}^\mathrm{H}[u]$ coincides with the expectation of $H_N/N$ on  $u^{\otimes N}$. Under the above assumptions on $V_\mathrm{DW}$ and $w$, the functional $\mathcal{E}^\mathrm{H}$ has a unique unit mass minimizer $u_+>0$ satisfying
\begin{equation}
\mathcal{E}^\mathrm{H}[u_+]=\min\Big\{ \mathcal{E}^\mathrm{H}[u]\;|\; u \in H^1(\mathbb{R}^d)\cap L^2( \mathbb{R}^d,V_\mathrm{DW}(x)dx),\; \|u\|_{L^2}=1 \Big\}.
\end{equation}

Through the Hartree minimizer $u_+$ we define the mean-field Hamiltonian
\begin{equation}
h_\mathrm{MF}:=-\Delta + V_{\mathrm{DW}} + \lambda w*|u_+|^2.
\end{equation}
By standard arguments one sees that $u_+$ is the ground state of the linear operator $h_{\mathrm{MF}}$, corresponding to an eigenvalue that we denote by $\mu_+$. Since $h_\mathrm{MF}$ has compact resolvent, its spectrum consists of eigenvalues only, and we thus write a spectral decomposition
\begin{equation} \label{eq:decomp_h}
h_\mathrm{MF}=\mu_+ \ket{u_+}\bra{u_+} + \mu_- \ket{u_-}\bra{u_-}+ \sum_{m\ge3}\mu_m \ket{u_m} \bra{u_m}.
\end{equation}
The second eigenvector $u_-$ is non-degenerate, and we can fix its phase so that $u_-(x)$ is positive for $x_1>0$ and negative for $x_1<0$. In fact, it is easy to see that $u_-$ must be odd with respect to reflections across the $\{x_1=0\}$ hyperplane (see \cite[Theorem 2.1 and Lemma 4.2]{OlgRou-20}).

The quantity capturing many spectral properties of $h_\mathrm{MF}$ is the tunneling parameter
\begin{equation}
T:= \exp\Big\{-\frac{2}{1+s/2}\big(L/2\big)^{1+s/2}\Big\},
\end{equation}
where $L$ is the distance between the two wells. Since we are working in a regime or large $L$, $T$ is to be understood as a small parameter. We remark that the quantity appearing in the exponent in $T$ is (twice) the Agmon distance within the potential $|x|^s$. It is know from the semiclassical analysis literature (see, e.g.,~\cite{Agmon-82,DimSjo-99,Helffer-88}) that this controls the absolute value of one-well eigenvectors in the region between the two potential wells, thus capturing the magnitude of the tunneling effect.
In our context, we proved in~\cite{OlgRou-20} that
\begin{equation} \label{eq:gap}
0< \mu_--\mu_+\simeq T, \qquad \mu_\mathrm{ex}-\mu_->C>0.
\end{equation}
The first bound shows that $T$ essentially coincides with the first gap of the mean-field Hamiltonian. The rest of the spectrum, in turn, is separated from $\mu_\pm$ by a uniform gap. These properties will have major consequences in our analysis, implying in particular that the two-dimensional subspace generated by $u_+$ and $u_-$ will be occupied to leading order, while the other modes contribute in a subleading way. It further follows from the results in \cite{OlgRou-20} that, within the space generated by $u_+$ and $u_-$, the linear combinations
\begin{equation} \label{eq:u_1u_2}
u_1:=\frac{u_++u_-}{\sqrt{2}},\qquad u_2:=\frac{u_+-u_-}{\sqrt{2}}
\end{equation}
are mostly localized, respectively, in the right and left wells. Notice that, in view of this, the identity
\begin{equation*}
 \mu_--\mu_+  = -2 \langle u_1,h_{\mathrm{MF}}\,u_2\rangle 
\end{equation*}
further vindicates the interpretation of $T \simeq \mu_--\mu_+$ as tunneling parameter.

Unless differently specified, from now on the notation
\begin{equation} \label{eq:basis}
\{ u_1,u_2,u_3,u_4\dots\}
\end{equation}
for an orthonormal basis of $L^2(\mathbb{R}^d)$ will be used to denote $u_1,u_2$ defined in \eqref{eq:u_1u_2} and $u_3,u_4,..$ defined by \eqref{eq:decomp_h}.

The decomposition of the spectrum in modes that (mostly) live in the left or right well holds in fact even for excited eigenvectors. One can suitably group the $u_m's$ for $m\ge3$ in pairs in such a way that
\begin{equation} \label{eq:right_left_modes}
u_{r,\alpha}:=\frac{u_{2\alpha+1}+u_{2\alpha+2}}{\sqrt{2}},\qquad u_{\ell,\alpha}:=\frac{u_{2\alpha+1}-u_{2\alpha+2}}{\sqrt{2}}
\end{equation}
for any $\alpha\ge1$ are mostly localized, respectively, in the right and left wells (see \cite[Theorem 2.2]{OlgRou-20}).

\subsection{Two-mode Hamiltonian}
To establish a link between the many-body Hamiltonian $H_N$ from \eqref{eq:H_N} and the two-mode Hamiltonian below, we rewrite $H_N$ using the standard bosonic creation and annihilation operators $a^*_m=a^*(u_m)$ and $a_m=a(u_m)$ as
\begin{equation} \label{eq:H_N_second_quant}
H_N= \sum_{m,n\ge1} h_{mn} a^*_m a_n+\frac{\lambda}{2(N-1)} \sum_{m,n,p,q\ge1} w_{mnpq} a^*_m a^*_n a_pa_q,
\end{equation}
with
\begin{equation}
\begin{split}
h_{mn}:=\;&\big\langle u_m,(-\Delta+V_\mathrm{DW})u_n \big\rangle\\
w_{mnpq}:=\;& \big\langle u_m\otimes u_n,w, u_p\otimes u_q\big\rangle.
\end{split}
\end{equation}
This expression is independent on the choice of the orthonormal basis $\{u_m\}$, but we consider here the special choice \eqref{eq:basis}.

In the bosonic mean-field analysis for bosons trapped by a fixed well~\cite{Seiringer-11,GreSei-13,LewNamSerSol-13} the leading order of the ground state energy $E(N)$ is completely captured by the Hartree minimizer $u_+$. The rest of the spectrum of $h_\mathrm{MF}$ is separated from $\mu_+$ by a finite gap, and the excited modes thus contribute to the $N$-body energy in a subleading way. In our double-well case, in turn, the gap between $\mu_-$ and $\mu_+$ is small. This means that we should expect an important contribution to the energy to come from particles occupying the mode $u_-$.

Motivated by this we define the orthogonal projection
\begin{equation} \label{eq:PQ}
P:=\ket{u_+}\bra{u_+}+ \ket{u_-}\bra{u_-}= \ket{u_1}\bra{u_1}+\ket{u_2}\bra{u_2},\qquad Q:= \mathbbm{1} - P= \sum_{m\ge3} \ket{u_m}\bra{u_m}
\end{equation}
and consequently the 2-mode Hamiltonian
\begin{equation} \label{eq:H_2-mode}
H_{2-\mathrm{mode}}:=P^{\otimes N} H_N P^{\otimes N}
\end{equation}
with ground state energy
\begin{equation}
 E_{2-\mathrm{mode}}:=\inf \Big\{ \left\langle \psi_N,H_N \psi_N\right\rangle,\;|\;\psi_N \in P^{\otimes N} \mathcal{H}_N,\;\int_{\mathbb{R}^{dN}}|\psi_N|^2=1 \Big\}.
\end{equation}
It is important to notice that $H_{2-\mathrm{mode}}$ is a $N$-body operator. This is in contrast to the Hartree functional, obtained by projecting $H_N$ on the span of a single $N$-body function. The 2-mode Hamiltonian has a complicated expression which we write explicitly in Proposition \ref{prop:H_2-mode}. In order to grasp its most important feature, we remark that, up to an additive constant and some subleading terms, $H_{2-\mathrm{mode}}$ coincides with the well-known Bose-Hubbard Hamiltonian
\begin{equation} \label{eq:BH}
H_\mathrm{BH}=\frac{\mu_+-\mu_-}{2}\big(a^*_1a_2+a^*_2a_1\big)+\frac{\lambda\, w_{1111}}{2(N-1)}\big(a^*_1 a^*_1 a_1a_1+a^*_2 a^*_2 a_2a_2\big).
\end{equation}
We will not address here the precise relation between $H_{2-\mathrm{mode}}$ and $H_\mathrm{BH}$, which the reader can find in \cite[Section 4]{OlgRouSpe-21}.

\subsection{Bogoliubov energy}

The bound \eqref{eq:heuristical_result} that we aim at proving will eventually follow by an energy expansion accurate up to terms that vanish as $N\to\infty$. This means that we will need to precisely compute the contributions to the ground state energy $E(N)$ that come from the one-body modes orthogonal to $u_1$ and $u_2$. These contribute to the so-called Bogoliubov energy, which we define in the following.

Let us consider the basis \eqref{eq:right_left_modes} of modes orthogonal to $u_1$ and $u_2$ and living (essentially) on the right or left well only. We correspondingly define
\begin{equation}
Q_r := \sum_{\alpha\ge1}\ket{u_{r,\alpha}}\bra{u_{r,\alpha}},\qquad Q_\ell := \sum_{\alpha\ge1}\ket{u_{\ell,\alpha}}\bra{u_{\ell,\alpha}}.
\end{equation}
These operators project, respectively, on the right- and left-well excited modes of $h_\mathrm{MF}$, and therefore orthogonally to $\mathrm{span}\{u_+,u_-\}=\mathrm{span}\{u_1,u_2\}$. We also define the operator trace with respect to these excited modes as
\begin{equation}
\mathrm{Tr}_{\perp,r}(A):= \sum_{\alpha \ge1}\langle u_{r,\alpha},A\, u_{r,\alpha}\rangle,\qquad \mathrm{Tr}_{\perp,\ell}(A):= \sum_{\alpha \ge1}\langle u_{\ell,\alpha},A\, u_{\ell,\alpha}\rangle
\end{equation}
and
\begin{equation}
\mathrm{Tr}_\perp (A):=\sum_{m\ge1} \langle u_m,A u_m\rangle.
\end{equation}
We also define two operators $D_r$ and $D_\ell$ by
\begin{equation}
D_r:=Q_r(h_\mathrm{MF}-\mu_+)Q_r,\qquad D_\ell:=Q_\ell(h_\mathrm{MF}-\mu_+)Q_\ell
\end{equation}
and two more operators $K_{11}$ and $K_{22}$ through their matrix elements
\begin{equation}
\begin{split}
\langle v, K_{11} u \rangle=\;&\frac{1}{2}\langle v\otimes u_1, w\, u_1\otimes u\rangle \\
\langle v, K_{22} u \rangle =\;& \frac{1}{2} \langle v \otimes u_2, w\, u_2\otimes u\rangle.
\end{split}
\end{equation}
We are finally ready to define the Bogoliubov energy
\begin{equation} \label{eq:E^bog}
\begin{split}
E^\mathrm{Bog}=\;&-\frac{1}{2}\mathrm{Tr}_{\perp,r}\left[ D_r+\lambda Q_r K_{11} Q_r-\sqrt{D_r^2+2\lambda D_r^{1/2}Q_rK_{11}Q_rD_r^{1/2} }\right]\\
&-\frac{1}{2}\mathrm{Tr}_{\perp,\ell}\left[ D_\ell+\lambda Q_\ell K_{22} Q_\ell-\sqrt{D_\ell^2+2\lambda D_\ell^{1/2}Q_\ell K_{22}Q_\ell D_\ell^{1/2} }\right].
\end{split}
\end{equation}
The two expressions in the right hand side, which coincide by reflection symmetry, represent the lowest eigenvalues to two independent (commuting) quadratic Bogoliubov Hamiltonians acting on the Fock space generated by right- or left-well modes only (compare for example with the expressions in \cite[Section 4 and 5]{GreSei-13}). It is worth noticing that, even though the summands inside each trace are not trace-class operators separately (in fact, $D_r$ and $D_\ell$ are unbounded), a non-trivial cancellation occurs, ensuring that the right hand side of \eqref{eq:E^bog} is well-defined (see e.g. \cite[Proposition 1]{GreSei-13})

\subsection{Main result}

Let us define the number operators for modes $u_1$, $u_2$, and for excited modes as the operators on $\mathcal{H}_N$
\begin{equation} \label{eq:cN}
\mathcal{N}_1= a^*_1 a_1,\qquad\mathcal{N}_2=a^*_2 a_2,\qquad\mathcal{N}_\perp =\sum_{m\ge3} a^*_m a_m= N-\mathcal{N}_1-\mathcal{N}_2.
\end{equation}
Our main result is the following theorem.

\begin{theorem} \label{thm:main}
	Assume that $T\sim N^{-\delta}$ for some $\delta>0$, and assume that $0<\lambda<\lambda_0$ for some $\lambda_0>0$ small enough. Then
	\begin{equation} \label{eq:main_variance}
	\lim_{N\to\infty} \frac{1}{N}\,\left\langle \left(\mathcal{N}_1-\mathcal{N}_2\right)^2\right\rangle_{\psi_\mathrm{gs}}=0
	\end{equation}
	and
	\begin{equation} \label{eq:main_energy}
	\lim_{N\to\infty}|E(N)-E_{2-\mathrm{mode}}-E^\mathrm{Bog}|=0.
	\end{equation}
\end{theorem}

The result \eqref{eq:main_variance} can be rephrased as
\begin{equation}
	\lim_{N\to\infty}\frac{1}{N} \,\left\langle \left(\mathcal{N}_1-\frac{N}{2}\right)^2\right\rangle_{\psi_\mathrm{gs}}=0,
\end{equation}
and the analogous one for $\mathcal{N}_2$, using the (non-trivial) Bose-Einstein condensation result (see \eqref{eq:a_priori_first})
\begin{equation}
\big\langle\big( N-\mathcal{N}_1-\mathcal{N}_2\big)\big\rangle_{\psi_\mathrm{gs}}\le C.
\end{equation}
By what we discussed previously in this section, \eqref{eq:main_variance} is in disagreement with a central limit theorem for fluctuations at a scale $\sqrt{N}$, therefore signaling the emergence of a correlated phase in the ground state.

The restriction on the magnitude of the coupling constant $\lambda$ is a technical assumption that we need for a very specific bound to hold (see Lemma \ref{lemma:coupling_constant}). We believe it could be relaxed.

The lack of a convergence rate for both \eqref{eq:main_variance} and \eqref{eq:main_energy} is a product of the fact that our derivation of the Bogoliubov energy involves a splitting into right and left wells all along the spectrum, with a bad uniformity in the tunneling parameter. We believe that significative improvements in our proof should allow to restate \eqref{eq:main_variance} in the form
\begin{equation*}
\frac{1}{N}\,\left\langle \left(\mathcal{N}_1-\mathcal{N}_2\right)^2\right\rangle_{\psi_\mathrm{gs}} \le C \max \{T^{1/2},{N}^{-1}\}.
\end{equation*}
The quantities on the right hand side are, at least heuristically, the two natural next contributions to the energy beyond $E_{2-\mathrm{mode}}$ and $E^\mathrm{Bog}$, $T^{1/2}$ already appearing as a remainder term in the energy \eqref{eq:H_2-mode_upper} of the 2-mode trial state, and $N^{-1}$ being the usual magnitude of third order contributions to the $N$-body energy in the mean-field regime~\cite{BosPetSei-21}.

The rest of the paper is structured as follows. Section \ref{sect:2_mode} contains a discussion of the 2-mode Hamiltonian and in particular upper and lower bounds on its minimal energy. Section \ref{sect:excitations} contains the main mathematical definitions that allow to properly rewrite the model in the excitation space, and the derivation of the Bogoliubov Hamiltonian as the operator governing the next-to-leading (beyond $H_{2-\mathrm{mode}}$) energy behavior of the system. In Section \ref{sect:diagonalization} we will discuss the minimization of the Bogoliubov Hamiltonian. Finally, in Section \ref{sect:proofs} we will show how the results presented imply the proof of Theorem \ref{thm:main}.

\section{The 2-mode Hamiltonian} \label{sect:2_mode}

The two mode Hamiltonian \eqref{eq:H_2-mode} describes a $N$-body system in which every variable is constrained to two one-body modes. Although an explicit formula for the associated eigenvalues and eigenvectors is unlikely to be obtainable, an amount of non-trivial information can still be deduced in the form of energy bounds.

Let us define the operator
\begin{equation*} \label{eq:T_operator}
\mathcal{T}:=\frac{\mu_+-\mu_-}{2}-\frac{\lambda}{N-1}w_{1112}\mathcal{N}_\perp-\frac{\lambda}{N-1}w_{1122}(\mathcal{N}_\perp-1)
\end{equation*}
and the constants
\begin{equation} \label{eq:constants}
\begin{split}
E_0 :=\;& Nh_{11} + \frac{\lambda N^2}{4(N-1)} (2 w_{1122} - w_{1212})\\
E_N^w:=\;&N\Big (\frac{\lambda N}{4(N-1)}(w_{1111}-4 w_{1122}+ 2 w_{1212})-\frac{\lambda}{2(N-1)}(w_{1111}+w_{1122})\Big)\\
\mu:=\;&h_{11}+\frac{\lambda}{2}w_{1111}+\frac{\lambda N}{2(N-1)}(w_{1212}-2w_{1122})-\frac{\lambda}{2(N-1)}w_{1122}\\
U:=\;&\frac{1}{4}(w_{1111}-w_{1212}).
\end{split}
\end{equation}
For a heuristic understanding of the above definitions, the reader should keep in mind that $\mathcal{N}_\perp\simeq C$ on states with few excitations, and therefore, by the bounds in \cite[Lemma 4.1]{OlgRouSpe-21} on the $w_{mnpq}$-coefficients, we should expect
\begin{equation*}
\mathcal{T} \simeq \frac{\mu_+-\mu_-}{2}.
\end{equation*}
Similarly, up to negligible contributions we have
\begin{equation*}
\mu\simeq \mu_+ \qquad\text{and}\qquad U \simeq \frac{w_{1111}}{4} \ge C >0,
\end{equation*}
and $w_{1111}$ is the typical Bose-Hubbard coupling constant for a quartic on-site term $a^*_1 a^*_1 a_1a_1$.

Using the above notations we have the identity
\begin{equation} \label{eq:H_2mode_identity}
\begin{split}
H_{2\mathrm{-mode}}=\;&E_0+E^w_N+\mathcal{T}\big(a^\dagger_1a_2+a^\dagger_2a_1\big)-\mu\mathcal{N}_\perp+\frac{\lambda U}{N-1}\left(\mathcal{N}_1-\mathcal{N}_2\right)^2\\
&+\frac{2\lambda}{N-1}w_{1122}\mathcal{N}^2_-+\frac{\lambda}{4(N-1)}(w_{1111}-2w_{1122}+w_{1212})\mathcal{N}_\perp^2.
\end{split}
\end{equation}
This formula was shown in~\cite[Proposition 4.2]{OlgRouSpe-21}. Its remarkable feature is that all the non-trivial interaction contributions (that is, all quartic terms in creators and annihilators), are either included in the variance operator $(\mathcal{N}_1-\mathcal{N}_2)^2$, with coupling constant (approximately) proportional to $ w_{1111}$ or in $\mathcal{N}_-^2$. The fact that the latter operator is positive and multiplied by the very small factor $N^{-1}w_{1122} \lesssim C T^2N^{-1}$ (see~\cite[eq. (4.6)]{OlgRouSpe-21}) allows to neglect its contribution both in the upper and lower bounds. To even better read \eqref{eq:H_2mode_identity} we remark that all $\mathcal{N}_\perp$ terms give a negligible contribution on states with few excitations, since they are multiplied by small factors, except for $-\mu\mathcal{N}_\perp$ which will contribute to the Bogoliubov energy. Moreover, the operator $\mathcal{T}\simeq \mu_+-\mu_-$ is the coupling constant for the hopping term $a^*_1 a_2+a^*_2 a_1$, once again justifying its interpretation as tunneling term. All this can be rephrased by saying that, up to additive constant terms and negligible contributions, the behavior of $H_{2-\mathrm{mode}}$ is well-captured by the Bose-Hubbard Hamiltonian
\begin{equation*}
H_\mathrm{BH}=\frac{\mu_+-\mu_-}{2}\big(a^*_1a_2+a^*_2a_1\big)+\frac{\lambda\, w_{1111}}{2(N-1)}\big(a^*_1 a^*_1 a_1a_1+a^*_2 a^*_2 a_2a_2\big)
\end{equation*}
already defined in \eqref{eq:BH}.

The above heuristic discussion is vindicated by the next result.

\begin{proposition}[\textbf{Upper and lower bounds for $H_{2-\mathrm{mode}}$}]\mbox{} \label{prop:H_2-mode}\\ 
	Assume that $T\sim N^{-\delta}$ for some $\delta>0$. Let $P$ be the projection oton the two-mode space defined in \eqref{eq:PQ}. Then there exists a normalized trial state $\psi_\mathrm{gauss}\in P^{\otimes N} \mathcal{H}_N$ such that, for every $\varepsilon>0$, there exists $C_\varepsilon>0$ such that
	\begin{equation} \label{eq:H_2-mode_upper}
	\langle \psi_\mathrm{gauss},H_{2-\mathrm{mode}} \psi_\mathrm{gauss}\rangle \le E_0+E^w_N+N \frac{\mu_+-\mu_-}{2}+ C_\varepsilon\max\{ T^{1/2-\varepsilon},N^{-1+\varepsilon\delta} \}.
	\end{equation}
	Moreover, we have the lower bound
	\begin{equation} \label{eq:H_2-mode_lower}
	H_{2-\mathrm{mode}}\ge E_0+E^w_N-\mu_+ \mathcal{N}_\perp + N \frac{\mu_+-\mu_-}{2}+\frac{\lambda U}{N-1}(\mathcal{N}_1-\mathcal{N}_2)^2-C_\varepsilon T^{1-\varepsilon}\mathcal{N}_\perp.
	\end{equation}
	Finally, we have
	\begin{equation} \label{eq:E_2_mode_bound}
	\begin{split}
	\Big| E_{2-\mathrm{mode}} - E_0-E^w_N-N\frac{\mu_+-\mu_-}{2}\Big|\le C_\varepsilon\max\{ T^{1/2-\varepsilon},N^{-1+\varepsilon\delta} \}.
	\end{split}
	\end{equation}
\end{proposition}

The proof of Proposition \ref{prop:H_2-mode} can be found in~\cite[Section 4]{OlgRouSpe-21}. For the lower bound, the most noteworthy point is the need to recognize that (the leading part of) $\mathcal{T}$ is the negative number $\mu_+-\mu_-$, and one can therefore use Cauchy-Schwartz's inequality on the hopping operator to show that $\mathcal{T}(a^*_1a_2+a^*_2a_1) \gtrsim N(\mu_+-\mu_-)/2$. The term proportional to $\mathcal{N}_-^2$ is instead positive and can thus be neglected. For the upper bound, in turn, the explicit trial state is
\begin{equation} \label{eq:psi_gauss}
\psi_\mathrm{gauss}:= \sum_{\substack{-\sigma_N^2 \le d\le \sigma_N^2\\ N+d\text{ even}}} c_d \,u_1^{\otimes(N+d)/2} \otimes_\mathrm{sym} u_2^{\otimes(N-d)/2},
\end{equation}
with gaussian weights
\begin{equation}
c_d:=\frac{1}{Z_N}e^{-d^2/4\sigma_N^2},\qquad|d|\le \sigma_N^2.
\end{equation}
Here $Z_N$ is a normalization factor, while the variance $\sigma_N^2$ is by definition
\begin{equation}
\sigma_N^2:=\begin{cases}
\sqrt{\mu_--\mu_+}\,N\qquad&\text{if }\delta>2\\
C \qquad&\text{otherwise}.
\end{cases}
\end{equation}

The function $\psi_\mathrm{gauss}$ manifestly lives in the $N$-body 2-mode space $P^{\otimes N} \mathcal{H}_N$, and it is constructed by linearly combining, with gaussian weights, configurations with different populations between right and left wells (i.e. different occupations of $u_1$ and $u_2$). This is 
not unlike a trial state for a fixed-well setting (that is, $L\simeq 1$ in our $V_\mathrm{DW}$), but the key difference is in the magnitude of the variance $\sigma_N^2$. For fixed wells the gap $\mu_--\mu_+$ is not small, and therefore $\sigma_N^2 \simeq N$. Expectations of observables on $\psi_\mathrm{gauss}$ would thus be satisfying a central limit theorem. In our double-well case, in turn, the smallness of $\mu_--\mu_+$ implies a reduction of the variance of $\psi_\mathrm{gauss}$. We believe $\psi_\mathrm{gauss}$ should capture the right order of magnitude of the variance in the $N$-body ground state in \eqref{eq:main_variance}, that is $\sqrt{\mu_--\mu_+}$. Unfortunately, larger error terms are created in the calculations of the Bogoliubov energy, and we are thus currently unable to obtain such a quantitative estimate.

\section{The excitation space and the Bogoliubov Hamiltonian} \label{sect:excitations}

In order to prove Theorem \ref{thm:main} we need to compute the $N$-body ground state energy $E(N)$ sharply up to terms that vanish as $N\to\infty$. We cannot expect this to be attained, for example as an upper bound, through a trial state of the form \eqref{eq:psi_gauss} that lives in the 2-mode space only. As happens in the one-well setting, excited modes give a contribution $O(1)$ to the energy, and Bogoliubov theory is the tool commonly used to characterize such a contribution. This is, roughly speaking, a way of treating the occupation of excited modes as a perturbation with respect to the full occupation of the Hartree minimizer. One shows that the overall contribution of excited modes to the $N$-body energy is described by the Bogoliubov Hamiltonian, a quadratic operator on the Fock space with base space $Q L^2(\mathbb{R}^d)$ (Fock space of excitations), with $Q$ defined in \eqref{eq:PQ}. Explicit diagonalization of the Bogoliubov Hamiltonian is typically an exactly solvable problem, and this allows to gain a very good control to the next-to-leading order properties of the full ground state problem. The above picture has been largely confirmed in the past fifteen years by a huge number of results on the rigorous justification of Bogoliubov theory, in different regimes, regularity assumptions, and space dimensions~\cite{Seiringer-11,GreSei-13,LewNamSerSol-13,DerNap-13,NamSei-14,BocBreCenSch-17,BocBreCenSch-19,FouSol-19,BasCenSch-21,NamTri-21,BreSchSch-21b}.

\subsection{The excitation space}
In order to apply the formalism of Bogoliubov theory to our setting we need to extend a number of definitions from~\cite{LewNamSerSol-13} to account for the presence of two low-energy modes that might be macroscopically occupied. We first notice that the decomposition
\begin{equation}
\begin{split}
\mathfrak{H}^N=\;&\Big(\mathrm{span}\{u_1\}\oplus\mathrm{span}\{u_2\}\oplus\bigoplus_{m\ge3}^\infty \mathrm{span}\{ u_m\}\Big)^{\otimes_{\mathrm{sym}} N}
\end{split}
\end{equation}
induces the existence of a unique way of writing every $\psi_N \in\mathcal{H}_N$ as
\begin{equation} \label{eq:wavefunction_expansion}
\begin{split}
\psi_N=\;&\sum_{s=0}^N\;\sum_{d=-N+s,\,-N+s+2,\,\dots}^{\dots,\,N-s-2,\,N-s} u_1^{\otimes (N-s+d)/2}\otimes_\mathrm{sym} u_2^{\otimes (N-s-d)/2}\otimes_\mathrm{sym} \Phi_{s,d}
\end{split}
\end{equation}
for some 
\begin{equation*}
\Phi_{s,d} \in \Big( \mathrm{span}\{u_1,u_2\}^\perp \Big)^{\otimes_{\mathrm{sym}}s}.
\end{equation*}
For each summand in \eqref{eq:wavefunction_expansion} the indexes $s$ and $d$ represent, respectively, the total number of excited particles and the difference between the occupation number of $u_1$ and $u_2$. For fixed $d$, the collection of $\{\Phi_{s,d}\}$ for $0\le s \le N$ belongs to the Fock space of excitations
\begin{equation}
\mathcal{F}_\perp :=\mathbb{C}\oplus\bigoplus_{s=1}^\infty\Big( \mathrm{span}\{u_1,u_2\}^\perp \Big)^{\otimes_{\mathrm{sym}}s}.
\end{equation}
The whole collection of $ \{\Phi_{s,d}\} $ for $0\le s \le N$ and $d=-N+s,-N+s+2,\dots,N-s-2,N-s$ belongs to the full space of excitations
\begin{equation}
\ell^2 (\mathcal{F}_\perp):= \bigoplus_{d\in\mathbb{Z}} \mathcal{F}_\perp.
\end{equation}

We define the generalization of the $U_N$ map from~\cite{LewNamSerSol-13} as the linear mapping associating to every $\psi_N$ its excitation content $\{\Phi_{s,d}\}$ according to the decomposition \eqref{eq:wavefunction_expansion}, that is,
\begin{equation}
U_N:\mathcal{H}_N \to \ell^2(\mathcal{F}_\perp),\qquad U_N \psi_N=\bigoplus_{s=0}^N \;\bigoplus_{d=-N+s,-N+s+2,\dots}^{N-s-2,N-s} \Phi_{s,d}.
\end{equation}
$U_N$ is a partial isometry, and it is actually unitary if the codomain is restricted from the whole $\ell^2(\mathcal{F}_\perp)$ to the sectors with $0\le s \le N$, $|d|\le N-s$ and $(N-s+2)/2\in\mathbb{N}$.

It is easy to check that conjugating products of the type $a^*_p a_q$ with $U_N$ has, to a good approximation with states on few excitations, the effect of replacing $a^*_1,a^*_2,a_1,a_2$ with $\sqrt{N/2}$ and leaving all other operators untouched. This suggest that $U_N$ is the correct operator to monitor the effect of perturbation beyond the occupation of $u_1$ and $u_2$ by exactly $N/2$ particles each. To state these facts more precisely we need to control the mapping to $\ell^2(\mathcal{F}_\perp)$ of operators on the $N$-body space $\mathcal{H}_N$. We define the operator
\begin{equation}
\mathfrak{D}:=U_N(\mathcal{N}_1-\mathcal{N}_2)U_N^* \qquad\text{which acts as}\qquad
(\mathfrak{D}\Phi)_{s,d}=d\Phi_{s,d}.
\end{equation}
Notice that $\mathfrak{D}^2$ is thus unitarily equivalent to the operator appearing in the expectation value \eqref{eq:main_variance}. We therefore refer to $\mathfrak{D}^2$, as well as to $(\mathcal{N}_1-\mathcal{N}_2)^2$, as the variance operator. We also define the shift operator
\begin{equation}
\Theta:\ell^2(\mathcal{F}_\perp)\to \ell^2(\mathcal{F}_\perp)\qquad\text{which acts as}\qquad (\Theta \Phi)_{s,d}=\Phi_{s,d-1}.
\end{equation}

In terms of these operators and the $\mathcal{N}_\perp$ defined in \eqref{eq:cN} we write the conjugation of operators of the type $a^*_p a_q$ by $U_N$ as (see~\cite[Lemma 3.6]{OlgRouSpe-21})
\begin{equation} \label{eq:conjugation}
\begin{split}
\mathcal{U}_Na^* _1a_1\mathcal{U}_N^*=\;&\frac{N-\mathcal{N}_\perp +\mathfrak{D}}{2}\\
\mathcal{U}_Na^* _1a_2\mathcal{U}_N^*=\;&\Theta\sqrt{\frac{N-\mathcal{N}_\perp +\mathfrak{D}+1}{2}}\sqrt{\frac{N-\mathcal{N}_\perp -\mathfrak{D}+1}{2}}\,\Theta\\
\mathcal{U}_Na^* _2a_2\mathcal{U}_N^*=\;&\frac{N-\mathcal{N}_\perp -\mathfrak{D}}{2}	\\
\mathcal{U}_Na^* _1a_m\mathcal{U}_N^*=\;&\Theta\sqrt{\frac{N-\mathcal{N}_\perp +\mathfrak{D}+1}{2}}\,a_m\\
\mathcal{U}_Na^* _2a_m\mathcal{U}_N^*=\;&\Theta^{-1}\sqrt{\frac{N-\mathcal{N}_\perp -\mathfrak{D}+1}{2}}\,a_m\\
\mathcal{U}_Na^{*}_ma_n \,\mathcal{U}_N =\;&a^{*}_ma_n^{}\\
\end{split}
\end{equation}
for all $m,n\ge3$.

\subsection{The Bogoliubov Hamiltonian}

Our first main goal is to conjugate $H_N-H_{2-\mathrm{mode}}$ with $U_N$. This procedure will give, up to suitably controlled remainders, the Bogoliubov Hamiltonian on $\ell^2(\mathcal{F}_\perp)$
\begin{equation} \label{eq:Bog_Hamiltonian}
\begin{split}
\mathbb{H}=\;&\sum_{m,n\ge3}\Big( -\Delta+V_\mathrm{DW}+\frac{\lambda}{2}w*|u_1|^2+\frac{\lambda}{2}w*|u_2|^2+{\lambda}K_{11}+{\lambda}K_{22}-\mu_+ \Big)_{mn}a^* _ma_n\\
&+\frac{\lambda}{2}\sum_{m,n\ge3}\big(K_{11} \big)_{mn}\Big( \Theta^{-2} a^* _ma^* _n+\Theta^2 a_ma_n\Big)\\
&+\frac{\lambda}{2}\sum_{m,n\ge3}\big( K_{22}\big)_{mn}\Big(  \Theta^2 a^* _ma^* _n+ \Theta^{-2}a_ma_n\Big)\\
&+\frac{\lambda}{2}\sum_{m,n\ge3} \big(K_{12}\big)_{mn} a^* _ma^* _n+\frac{\lambda}{2} \sum_{m,n\ge3}\big(K_{12}^{*}\big)_{mn} a_ma_n\\
&+\frac{\lambda}{2}\sum_{m,n\ge3}\big(K_{12} +w*(u_1u_2)\big)_{mn}\Theta^2 a^* _ma_n+\frac{\lambda}{2}\sum_{m,n\ge3}\big(K_{12}^\ast +w*(u_1u_2)\big)_{mn}\Theta^{-2} a^* _ma_n
\end{split}
\end{equation}
where the operators $K_{ij}$ are defined by
\begin{equation}
\begin{split}
\langle v,K_{11}u\rangle=\;&\frac{1}{2}\langle v\otimes u_1\,,\, w \,u_1\otimes u\rangle\\
\langle v,K_{22}u\rangle=\;&\frac{1}{2}\langle v\otimes u_2\,,\,w\,u_2\otimes u\rangle\\[2mm]
\langle v,K_{12}u\rangle=\;&\langle v\otimes u_1\,,\,w\,u_2\otimes u\rangle.
\end{split}
\end{equation}
The emergence of $\mathbb{H}$ is precisely stated in the following result.

\begin{proposition}[\textbf{Derivation of the Bogoliubov Hamiltonian}] \label{prop:bogoliubov}\mbox{}\\
	Let $\Phi\in\ell^2(\mathfrak{F}_\perp)$ be of the form $\Phi = U_N \psi_N$ for some $\psi_N \in \mathcal{H}_N$ satisfying the bounds
	\begin{equation} \label{eq:bounds_test_state}
	\begin{split}
	\big\langle \Phi,(\mathcal{N}_\perp^2+1)\Phi\big\rangle \le\;& C\\
	\langle \Phi, \mathfrak{D}^2 \Phi\rangle \le\;& CN\\
	\big\langle U_N^* \Phi,\,\mathcal{N}_-\, U_N^*\Phi\big\rangle \le\;& C \min\{ N,\frac{1}{T}\}
	\end{split}
	\end{equation}

	 Then
	\begin{equation} \label{eq:derivation_bogoliubov}
	\begin{split}
	\lim_{N\to\infty}\bigg|\Big\langle\Phi,\Big( {U}_N(&H_N-H_{2\mathrm{-mode}}){U}_N^*- \mathbb{H}- \mu_+ \mathcal{N}_\perp\Big)\Phi\Big\rangle\\
	&-\frac{\lambda}{\sqrt{2(N-1)}}\sum_{m\ge3}\Big\langle \Phi, \Big(\big(w_{+1-m}\,\Theta+w_{+2-m}\, \Theta^{-1}\big) a_m\mathfrak{D}+\mathrm{h.c.}\Big)\Phi\big\rangle\bigg|=0.
	\end{split}
	\end{equation}
\end{proposition}

The limit in \eqref{eq:derivation_bogoliubov} could in fact be stated in terms of an explicit decay rate that depends on the expectations in \eqref{eq:bounds_test_state}, as in (see \cite[Proposition 5.1]{OlgRouSpe-21})
\begin{equation*}
\begin{split}
\bigg|\Big\langle\Phi,\Big( {U}_N(&H_N-H_{2\mathrm{-mode}}){U}_N^*- \mathbb{H}- \mu_+ \mathcal{N}_\perp\Big)\Phi\Big\rangle\\
&-\frac{\lambda}{\sqrt{2(N-1)}}\sum_{m\ge3}\Big\langle \Phi, \Big(\big(w_{+1-m}\,\Theta+w_{+2-m}\, \Theta^{-1}\big) a_m\mathfrak{D}+\mathrm{h.c.}\Big)\Phi\big\rangle\bigg|\\
\le\;&\frac{C}{N^{1/4}}\left( \left\langle\Phi, (\mathcal{N}_\perp^2+1)\Phi\right\rangle+\frac{1}{N} \left\langle \Phi, \mathfrak{D}^2 \Phi\right\rangle \right)+C_\varepsilon \frac{T^{1-\varepsilon}}{N^{1/4}}\left\langle U_N^* \Phi, \mathcal{N}_- U_N^* \Phi\right\rangle^{3/4} \left\langle \Phi,\mathcal{N}_\perp^2\Phi\right\rangle^{1/4}.
\end{split}
\end{equation*}
Notice that the expectation values in \eqref{eq:bounds_test_state} are tailored to what the a priori bounds on the ground state in Proposition \ref{prop:a_priori} will be. The interpretation of the bound on the expectation of $\mathcal{N}_-$ will thus be described, in the case of $U_N^*\Phi=\psi_\mathrm{gs}$, after Proposition \ref{prop:a_priori}.

The proof of Proposition \ref{prop:bogoliubov} can be found in \cite[Section 5]{OlgRouSpe-21}. It follows in part the proof of \cite[Proposition 5.1]{LewNamSerSol-13}, that is, the separation of the various terms of $H_N$ written in the second quantized form \eqref{eq:H_N_second_quant} depending on the number of operators $a_m,a^*_m$ with $m\ge3$ they contain, and their careful analysis with the aim of either extracting the important contributions of proving that they are negligible. It turns out that:
\begin{itemize}
	\item cubic and quartic terms in terms of $a_m,a^*_m$ for $m\ge3$ are uniformly small on states satisfying the first of the \eqref{eq:bounds_test_state};
	\item quadratic terms yield $\mathbb{H}$ up to remainders that are controllable on states satisfying the first two of the \eqref{eq:bounds_test_state}.
\end{itemize}

An important difference with respect to the single-well setting is however the appearance in \eqref{eq:derivation_bogoliubov} of linear terms containing one single $a_m,a^*_m$. This is one of the main novel technical aspect of our result in~\cite{OlgRouSpe-21}. In the standard proofs of Bogoliubov theory, linear terms are negligible thanks to an important cancellation occurring since the low-energy mode beyond which one expands is the exact Hartree minimizer. Such a cancellation occurs in our setting as well (see \cite[Proposition 5.4]{OlgRouSpe-21}), but the fact that our low-energy space is two-dimensional induces the appearance of a number of extra linear terms. It turns out that some of them are small due to a further cancellation occurring thanks to the optimality of the choice of $u_1$ and $u_2$ within the subspace they generate. The linear terms in \eqref{eq:derivation_bogoliubov}, in turn, are not a priori negligible. In fact, it can easily be shown that, as expectation on the ground state,
\begin{equation*}
\begin{split}
\frac{\lambda}{\sqrt{2(N-1)}}\bigg|\sum_{m\ge3}\Big\langle U_N &\psi_\mathrm{gs}, \Big(\big(w_{+1-m}\,\Theta+w_{+2-m}\, \Theta^{-1}\big) a_m\mathfrak{D}+\mathrm{h.c.}\Big)U_N\psi_\mathrm{gs}\big\rangle\bigg| \\
\le\;& C \langle \mathcal{N}_\perp \rangle_{\psi_\mathrm{gs}}^{1/2}\,\Big\langle \frac{1}{N}\big(\mathcal{N}_1-\mathcal{N}_2\big) \Big\rangle_{\psi_\mathrm{gs}}^{1/2}.
\end{split}
\end{equation*}
Since the expectation of $\mathcal{N}_\perp$ on $\psi_\mathrm{gs}$ is of order $O(1)$ (see \eqref{eq:a_priori_first} below), we see that the linear terms in \eqref{eq:derivation_bogoliubov} are small if the main result \eqref{eq:main_variance} holds true. But this is precisely what we want to prove at the very end, so the argument cannot hold. Notice that we could in principle control the linear terms as a further lower bound by
\begin{equation*}
-\frac{C\varepsilon}{N}(\mathcal{N}_1-\mathcal{N}_2)^2- \frac{C}{\varepsilon} \mathcal{N}_\perp.
\end{equation*}
For $\varepsilon$ small enough the variance term can be absorbed by the positive variance term in the 2-mode Hamiltonian \eqref{eq:H_2-mode_lower}. On the other hand, the remainder term $\sim \varepsilon^{-1} \mathcal{N}_\perp$ would spoil the exact calculation of the Bogoliubov energy, thus preventing to prove any of the main results in Theorem \ref{thm:main}. For these reasons we have to keep track of the extra linear terms all through the proof, and show that their presence does not have an important influence.

\subsection{A priori bounds on $\psi_\mathrm{gs}$}

The results presented so far are enough to compute the ground state energy only up to an energy that differs from the true ground state energy $E_N$ by a fixed constant. The reason for this is that the sharp $O(1)$ contribution to the energy can only be computed by minimizing the Bogoliubov Hamiltonian. This is certainly needed in order to prove Theorem \ref{thm:main}, and it will be the content of the next section. Nonetheless, the above results allow to deduce an amount of non-trivial bounds on the ground state $\psi_\mathrm{gs}$. Before stating them, we introduce the notation
\begin{equation}
\mathrm{d}\Gamma_\perp (A):=\sum_{m,n\ge3}\langle u_m,A u_n\rangle a^*_m a_n
\end{equation}
for the second quantization, restricted to the excited modes, of a self-adjoint operator $A$ on $L^2(\mathbb{R}^d)$.

\begin{proposition}[\textbf{A priori bounds}]\mbox{} \label{prop:a_priori} \\
	The following estimates hold.
	\begin{itemize}
		\item \textbf{Number and energy of excitations.}
		\begin{equation} \label{eq:a_priori_first}
		\begin{split}
		\langle	\mathcal{N}_\perp\rangle_{\psi_\mathrm{gs}} \le\;& C\\
		\langle \mathrm{d}\Gamma_\perp(h_\mathrm{MF}-\mu_+)\rangle_{\psi_\mathrm{gs}} \le\;&C \\
		\langle \mathcal{N}_-\rangle_{\psi_\mathrm{gs}} \le\;& C_\varepsilon\min\{ N,T^{-1-\varepsilon}\}.
		\end{split}
		\end{equation}
		\item \textbf{Variance.}
		\begin{equation} \label{eq:a_priori_variance}
		\big \langle \big(\mathcal{N}_1-\mathcal{N}_2\big)^2\big\rangle_{\psi_\mathrm{gs}} \le CN.
		\end{equation}
		\item \textbf{Second moments.}
		\begin{equation} \label{eq:a_priori_second}
		\big\langle \mathcal{N}_\perp^2\big\rangle_{\psi_\mathrm{gs}}+\big\langle \mathrm{d}\Gamma_\perp(h_\mathrm{MF}-\mu_+)\mathcal{N}_\perp\big\rangle_{\psi_\mathrm{gs}} \le C.
		\end{equation}
	\end{itemize}
\end{proposition}

The first two bounds in \eqref{eq:a_priori_first} and the bound \eqref{eq:a_priori_second} are BEC-type results: the number and energy of particles occupying excited modes in the ground state remains bounded in the large $N$ limit. The result \eqref{eq:a_priori_variance} states that the scaling behaviour of the variance is not worse that the one in the single-well case. We will use this to control error terms in a bootstrap-type argument that will eventually prove Theorem \ref{thm:main}. The third bound in \eqref{eq:a_priori_first} deserves some discussion. While the estimate in terms of $N$ is trivial, the $T^{-1-\varepsilon}$ has the physical interpretation of (approximately) the inverse of the gap between $\mu_+$ and $\mu_-$ (see \eqref{eq:gap}), and it should thus indeed control the occupation of $\mathcal{N}_-$. In our assumption $T\sim N^{-\delta}$ for some $\delta>0$, and therefore the bound in terms of $T^{-1-\varepsilon}$ is worse than the one in terms of $N$ when $\delta>1$. Nonetheless, since all error terms involving $\mathcal{N}_-$ will be multiplied by small coefficients proportional to positive powers of $T$ (see for example the $\mathcal{N}_-^2$ term in \eqref{eq:H_2mode_identity}), this estimate will suffice to our purposes.

The proof can be found in \cite[Section 6]{OlgRouSpe-21}. The overall strategy is similar to the proof of analogous results for the one-well setting, and indeed some arguments are borrowed from \cite{GreSei-13}. The more intricated nature of the double-well setting, however, forces a deviation from those proofs. In particular, \eqref{eq:a_priori_variance} and \eqref{eq:a_priori_second} cannot proven separately in a direct way. We first prove an intermediate version of \eqref{eq:a_priori_second} in which the variance operator appears in the right hand side. This allows to deduce \eqref{eq:a_priori_variance}, while \eqref{eq:a_priori_second} in its sharp form is later deduced through a bootstrap argument.

\section{Splitting the Bogoliubov Hamiltonian: lower bound on the Bogoliubov energy} \label{sect:diagonalization}
 
The main message of Proposition \ref{prop:bogoliubov} was that, once the 2-mode Hamiltonian is subtracted from $H_N$, the main contribution to the energy comes from the quadratic Bogoliubov Hamiltonian, up to some a priori non-negligible linear terms, i.e.
\begin{equation} \label{eq:overall_excitation_energy}
\mathbb{H}+\frac{\lambda}{\sqrt{2(N-1)}}\sum_{m\ge3} \Big(\big(w_{+1-m}\,\Theta+w_{+2-m}\, \Theta^{-1}\big) a_m\mathfrak{D}+\mathrm{h.c.}\Big).
\end{equation}
We show in the present section how to find an efficient energy lower bound for this operator.

Our overall strategy is to define new creation and annihilation operators that coincide with the original ones up to a translation (shift) tuned in order to absorb the linear terms. This amounts to a square-completion-type argument in \eqref{eq:overall_excitation_energy}, the effects of which are:
\begin{itemize}
	\item $\mathbb{H}$ is replaced by a new $\widetilde{\mathbb{H}}$ which is still a quadratic Hamiltonian but in terms of shifted creation and annihilation operators;
	\item a negative term depending on no $a_m,a_m^*$ appears to correct for the square completion;
	\item the technical trade-off for this is that the new creators and annihilators will satisfy perturbed canonical commutation relations
\end{itemize}

Unfortunately, the direct application of the above strategy runs into technical difficulties. Due to the presence of the operators $\Theta$ and $\mathfrak{D}$ in $\mathbb{H}$ and in the linear terms, the coefficients of the Bogoliubov Hamiltonian (that is, everything that multiplies $a_m,a_m^*$) have a non-trivial commutator with the shifted creators and annihilators. This makes an explicit diagonalization of $\mathbb{H}_\mathrm{shift}$ quite hard from the technical point of view.

To circumvent these problems we adopt a partly different approach to the problem of minimizing \eqref{eq:overall_excitation_energy}. This will be the content of the present section.

\subsection{Reduction to right and left modes}

Let us recall the two bases of the excitation space $\{u_m\}_{m \ge3}$ and $\{u_{r,\alpha},u_{\ell,\alpha}\}_{\alpha \ge 1}$ from, respectively, \eqref{eq:basis} and \eqref{eq:right_left_modes}. We will write a decomposition of \eqref{eq:overall_excitation_energy} into a part which only couples modes of type $u_{r,\alpha}$ among themselves, and a part which only couples modes of type $u_{\ell,\alpha}$ among themselves. All other terms, that is all those terms in \eqref{eq:overall_excitation_energy} that couple modes living in different wells, should be expected at least heuristically to be small in our regime of small tunneling. Unfortunately, since \eqref{eq:overall_excitation_energy} involves the whole spectral decomposition of $h_\mathrm{MF}$ up to arbitrarily large eigenvalues, it is hard to precisely control the tunneling of modes whose energy is far above the energy barrier between the two wells. To this end we will introduce a cutoffin the spectrum of $h_\mathrm{MF}$, and remove it only at the very end of the proofs, after the $N,L\to\infty$ limits.

We define the orthogonal projections
\begin{equation} \label{eq-def_P_<M}
Q_{\le M}:= \sum_{1\le \alpha \le M}\Big( \ket{u_{2\alpha+1}}\bra{u_{2\alpha+1}}+\ket{u_{2\alpha+2}}\bra{u_{2\alpha+2}} \Big)=\sum_{1\le \alpha \le M}\Big( \ket{u_{r,\alpha}}\bra{u_{r,\alpha}}+\ket{u_{\ell,\alpha}}\bra{u_{\ell,\alpha}} \Big)
\end{equation}
and
\begin{equation*}
Q_{> M}:=\sum_{\alpha > M}\Big( \ket{u_{2\alpha+1}}\bra{u_{2\alpha+1}}+\ket{u_{2\alpha+2}}\bra{u_{2\alpha+2}} \Big)=\mathbbm{1}-Q_{\le M}-\ket{u_+}\bra{u_+}-\ket{u_-}\bra{u_-}.
\end{equation*}

We also define the $\Theta$-shifted creation annihilation operators for modes living on the right or left wells as
\begin{equation}
b_{r,\alpha}:=\Theta \,a(u_{r,\alpha})\quad b_{r,\alpha}^*:=\Theta^{-1}a^*(u_{r,\alpha})\qquad
c_{\ell,\alpha}:= \Theta^{-1}a(u_{\ell,\alpha})\quad c^*_{\ell,\alpha}:=\Theta a^*(u_{\ell,\alpha}).
\end{equation}
These satisfy the following straightforwardly verifiable commutation relations
\begin{equation}
\begin{split}
[b_{r,\alpha},b^*_{r,\beta}]=[b_{r,\alpha},c_{\ell,\beta}]=[b_{r,\alpha},c^*_{\ell,\beta}]=\;&0\\
[b_{r,\alpha},b^*_{r,\beta}]=[c_{\ell,\alpha},c^*_{\ell,\beta}]=\;&\delta_{\alpha,\beta}.
\end{split}
\end{equation}

Through the above operators we define the right- and left-well Bogoliubov Hamiltonians with cutoff $M$
\begin{equation} \label{eq:H_bog_cutoff}
\begin{split} 
\mathbb{H}_\mathrm{right}^{(M)}:=\;&\sum_{1\le \alpha,\beta\le M}\left\langle u_{r,\alpha},\Big(h_\mathrm{MF}-\mu_++{\lambda}K_{11}\Big) u_{r,\beta}\right\rangle  b^\dagger _{r,\alpha} b_{r,\beta}\\
&+\frac{\lambda}{2}\sum_{1\le \alpha,\beta\le M}\left\langle u_{r,\alpha}, K_{11} u_{r,\beta}\right\rangle \left( b^\dagger _{r,\alpha} b^\dagger _{r,\beta}+ b_{r,\alpha} b_{r,\beta}\right)\\
\mathbb{H}_\mathrm{left}^{(M)}:=\;&\sum_{1\le \alpha,\beta\le M}\left\langle u_{\ell,\alpha},\Big(h_\mathrm{MF}-\mu_++{\lambda}K_{22}\Big) u_{\ell,\beta}\right\rangle  c^\dagger _{\ell,\alpha} c_{\ell,\beta}\\
&+\frac{\lambda}{2}\sum_{1\le \alpha,\beta\le M}\left\langle u_{\ell,\alpha}, K_{22} u_{\ell,\beta}\right\rangle \left( c^\dagger _{\ell,\alpha} c^\dagger _{\ell,\beta}+ c_{\ell,\alpha} c_{\ell,\beta}\right)\,,
\end{split}
\end{equation}
These are obtained by rewriting $\mathbb{H}$ in terms of the basis $\{u_{r,\alpha},u_{\ell,\alpha}\}_{\alpha\ge1}$ and neglecting all terms in which both a mode $u_{r,\alpha}$ and a mode $u_{\ell,\beta}$ appear, as well as all modes with index larger than $M$. It turns out that, to leading order, only $\mathbb{H}_\mathrm{right}^{(M)}$ and $\mathbb{H}^{(M)}_\mathrm{left}$ contribute to the minimization of $\mathbb{H}$, as shown by the next result.

\begin{proposition}[\textbf{Reduction to right and left modes}] \mbox{} \label{prop:reduction}\\
	Assume that $\Phi \in\ell^2(\mathcal{F}_\perp)$ satisfies
	\begin{equation}
	\left\langle \mathrm{d}\Gamma_\perp(h_\mathrm{MF}-\mu_+)(\mathcal{N}_\perp+1)+\frac{\mathfrak{D}^2}{N}+\mathcal{N}_\perp^2+1 \right\rangle \le C
	\end{equation}
	for a constant $C$ that does not depend on $N$. For every energy cutoff $\Lambda$, let $M_\Lambda$ be the largest integer such that $\mu_{2M_\Lambda+2} \le \Lambda$, where $\{\mu_m\}_{m\ge3}$ are the excited eigenvalues of $h_\mathrm{MF}$ in increasing order.
	\begin{itemize}
		\item \textbf{Bogoliubov Hamiltonian.} We have
		\begin{equation}
		\begin{split} \label{eq:reduction_l_r}
		\bigg| \Big\langle \mathbb{H}-\mathbb{H}_\mathrm{right}^{(M_\Lambda)}&-\mathbb{H}_\mathrm{left}^{(M_\Lambda)}-\mathrm{d}\Gamma_\perp\left(Q_{> M_\Lambda} \left(h_{\mathrm{MF}}-\mu_+\right)Q_{> M_\Lambda}\right) \Big\rangle_\Phi\bigg|\\
		\le\;& C_\Lambda o_N(1)+\frac{C}{\left(\mu_{2M_\Lambda+2}-\mu_+ \right)^{1/2}}
		\end{split}
		\end{equation}
		for two constants $C_\Lambda,C$ that do not depend on $N$.
		\item \textbf{Linear terms} We have
		\begin{equation}
		\begin{split} \label{eq:reduction_linear}
		 \Bigg| &\frac{\lambda}{\sqrt{2(N-1)}}\sum_{m\ge3} \Big\langle\big(w_{+1-m}\,\Theta+w_{+2-m}\, \Theta^{-1}\big) a_m\mathfrak{D}+\mathrm{h.c.}\Big\rangle_{\Phi} \\
		&\;\;-\frac{\lambda}{\sqrt{2(N-1)}}\sum_{1\le \alpha \le M_\Lambda} \big\langle u_1, w*(u_+u_-) u_{r,\alpha}\big\rangle\,\big\langle b_{r,\alpha}\mathfrak{D}+\mathrm{h.c.}\big\rangle_\Phi\\
		&\;\;-\frac{\lambda}{\sqrt{2(N-1)}}\sum_{1\le \alpha \le M_\Lambda} \big\langle u_2, w*(u_+u_-) u_{\ell,\alpha}\big\rangle\,\big\langle c_{\ell,\alpha}\mathfrak{D}+\mathrm{h.c.}\big\rangle_\Phi\Bigg| \\
		&\qquad\qquad\le\; C_\Lambda o_N(1)+\frac{C}{(\mu_{2M_\Lambda+2}-\mu_+)}^{1/2}.
		\end{split}
		\end{equation}
	\end{itemize}
\end{proposition}

Proposition \ref{prop:reduction} is proven in \cite[Sections 5.4 and 5.5]{OlgRouSpe-21}. The proof is based on a number of estimates on the tunneling terms (terms coupling right well modes with left well modes) that follow from the results in~\cite{OlgRou-20}.

It is worth mentioning that the $\mathrm{d}\Gamma_\perp$ operator appearing in \eqref{eq:reduction_l_r} is positive, and it can thus be discarded for a lower bound. In an upper bound it would cause more troubles since it yields a finite contribution as $N\to\infty$ (it only vanishes in the infinite cutoff limit $\Lambda\to\infty$). The splitting into right and left modes with cutoff will however not be needed when computing the energy of the trial state. Notice also that \eqref{eq:reduction_linear} shows that, just like $\mathbb{H}$, the non-negligible linear terms split into two parts, with an error that we can control in the large $N$ limit followed by the large cutoff limit.

\subsection{Lower bound for the Bogoliubov energy}

The results of Proposition \ref{prop:reduction} show that, for a lower bound on \eqref{eq:overall_excitation_energy}, it is enough to reduce ourselves to the sum of the two operators (neglecting the positive $\mathrm{d}\Gamma_\perp$ term in \eqref{eq:reduction_l_r})
\begin{equation*}
\begin{split}
\mathbb{H}_{\mathrm{right,shift}}^{(M)}:=\;&\mathbb{H}_\mathrm{right}^{(M)}+\frac{\lambda}{\sqrt{2(N-1)}}\sum_{1\le \alpha \le M} \big\langle u_1, w*(u_+u_-) u_{r,\alpha}\big\rangle\,\big( b_{r,\alpha}\mathfrak{D}+\mathrm{h.c.}\big)\\
\mathbb{H}_{\mathrm{left,shift}}^{(M)}:=\;&\mathbb{H}_\mathrm{left}^{(M)}+\frac{\lambda}{\sqrt{2(N-1)}}\sum_{1\le \alpha \le M} \big\langle u_2, w*(u_+u_-) u_{\ell,\alpha}\big\rangle\,\big( c_{\ell,\alpha}\mathfrak{D}+\mathrm{h.c.}\big)
\end{split}
\end{equation*}
for large enough $M$. As was the case for \eqref{eq:overall_excitation_energy}, this is again the sum of a quadratic and a linear expression in terms of creators and annihilators for excited modes. The fact that the two operators involve creators and annihilators of different modes (the $u_{r,\alpha}$ and $u_{\ell,\beta}$ from \eqref{eq:right_left_modes}), however, makes the minimization problem significantly simpler. We will now be able to apply the square completion argument sketched at the beginning of the present section.

Let us introduce the orthogonal projections
\begin{equation}
\begin{split}
Q_{r,\le M}:=\;&Q_r \,Q_{\le M}=\sum_{1\le \alpha \le M} \ket{u_{r,\alpha}}\bra{u_{r,\alpha}}\\
Q_{\ell,\le M}:=\;&Q_\ell\, Q_{\le M}=\sum_{1\le \alpha \le M} \ket{u_{\ell,\alpha}}\bra{u_{\ell,\alpha}}.
\end{split}
\end{equation}

\begin{proposition}[\textbf{Lower bound for shifted Hamiltonians}] \mbox{}\label{prop:lower_shifted} \\
	For any $\Phi\in\ell^2(\mathcal{F}_\perp)$ we have
	\begin{equation} \label{eq:lower_shifted}
	\begin{split}
	\big\langle \mathbb{H}_\mathrm{right,shift}^{(M)}&\big\rangle_{\Phi}+\big\langle \mathbb{H}_\mathrm{right,shift}^{(M)}\big\rangle_{\Phi} \\
	\ge\;& E^\mathrm{Bog}-\frac{\lambda^2}{2(N-1)} \Big(\big\langle u_1,K_{11}W_{r,\le M} K_{11} u_1\big\rangle+\big\langle u_2,K_{22}W_{\ell,\le M} K_{22} u_2\big\rangle\Big) \langle \mathfrak{D}^2\rangle_{\Phi}\\
	&-\frac{C_M}{\sqrt{N}}\langle\mathcal{N}_\perp^2+1\rangle_{\Phi}-\frac{C_\varepsilon T^{1/2-\varepsilon}}{N}\langle \mathfrak{D}^2\rangle_{\Phi},
	\end{split}
	\end{equation}
	where
	\begin{equation}
	\begin{split}
	W_{r,\le M}:=\;&Q_{r,\le M}\big( Q_{r,\le M}(h_\mathrm{MF}-\mu_++2\lambda K_{11})Q_{r,\le M} \big)^{-1}Q_{r,\le M}\\
	W_{\ell,\le M}:=\;&Q_{\ell,\le M}\big( Q_{\ell,\le M}(h_\mathrm{MF}-\mu_++2\lambda K_{22})Q_{\ell,\le M} \big)^{-1}Q_{\ell,\le M}.
	\end{split}
	\end{equation}
\end{proposition}

The proof of Proposition \ref{prop:lower_shifted} can be found in \cite[Section 7]{OlgRouSpe-21}. The overall strategy is, as already anticipated, a square-completion argument to rewrite the left hand side of \eqref{eq:lower_shifted}. This produces the negative terms containing $W_{r,\le M}$ and $W_{\ell,\le M}$ that appear in the right hand side of \eqref{eq:lower_shifted}, together with new quadratic Hamiltonians in terms of shifted creation and annihilation operators. The minimization of the latter Hamiltonians eventually produce the quantity $E^\mathrm{Bog}$.

The minimization (in fact, the full diagonalization) of bosonic quadratic Hamiltonians on Fock space is a well-understood problem, for which we refer, for example, to \cite{GreSei-13,LewNamSerSol-13,Derezinski-17}. In the standard setting it is possible to give rather explicit formulas for eigenvalues and eigenvectors of quadratic Hamiltonians. This is however one of the points in which our proofs have to be drastically adapted with respect to the one-well setting. The reason is that, after the square completion, the new creation and annihilation operators satisfy non-standard commutation relations (see \cite[Eq. (7.4)]{OlgRouSpe-21}). This prevents us from, among other things, directly using known theorems of diagonalization of quadratic Hamiltonians through unitary operators implementing Bogoliubov transformations. Rather, we adapt to our setting the explicit symplectic transformation mixing creators and annihilators that was already used in \cite{GreSei-13}. In our case this produces extra remainder terms due to the modification in the canonical commutation relations. We are however able to control them using the fact that the cutoff reduces the problem to a finite number of modes, and the fact that we are only looking for a lower bound.

As a last result of this section we state the next lemma, which is the only point in which the smallness assumption on $\lambda$ in Theorem \ref{thm:main} plays a role. Let us first notice that the operator multiplying the negative $W_{r,\le M}$ and $W_{\ell,\le M}$ terms is the variance
\begin{equation*}
\mathfrak{D}^2 = U_N (\mathcal{N}_1-\mathcal{N}_2)^2 U_N^*.
\end{equation*} 
It thus makes sense to collect and compare the coefficients of the operator $(N-1)^{-1}(\mathcal{N}_1-\mathcal{N}_2)^2$ inside the full lower bound for $H_N$. These are
\begin{equation}
\lambda U-\frac{\lambda^2}{2}\Big(\big\langle u_1,K_{11}W_{r,\le M} K_{11} u_1\big\rangle+\big\langle u_2,K_{22}W_{\ell,\le M} K_{22} u_2\big\rangle\Big),
\end{equation}
where $U$ was defined in \eqref{eq:constants} and appears in the lower bound for $H_{2-\mathrm{mode}}$ \eqref{eq:H_2-mode_lower}.

\begin{lemma}[\textbf{Variance coefficients}] \mbox{} \label{lemma:coupling_constant}\\
	Let us assume that $0<\lambda\le\lambda_0$ for some $\lambda_0>0$ small enough. Then
	\begin{equation} \label{eq:lower_bound_coefficient}
	\lambda U-\frac{\lambda^2}{2}\Big(\big\langle u_1,K_{11}W_{r,\le M} K_{11} u_1\big\rangle+\big\langle u_2,K_{22}W_{\ell,\le M} K_{22} u_2\big\rangle\Big)  \ge c \lambda
	\end{equation}
	for some $c>0$.
\end{lemma}
The proof can be found in \cite[Lemma 8.5]{OlgRouSpe-21}. It is a straightforward consequence of the fact that
\begin{equation*}
  K_{11}W_{r,\le M} K_{11} +K_{22}W_{\ell,\le M} K_{22}\le C
\end{equation*}
uniformly in $N,M,\lambda$. A better upper bound on this operator would possibly allow for a more efficient way of obtaining a strictly positive quantity on the right hand side of \eqref{eq:lower_bound_coefficient}

\section{Proof of the main result} \label{sect:proofs}

We now have all the intermediate results that are needed in order to prove Theorem \ref{thm:main}. The proof of both statements follows by matching upper and lower bounds to the ground state energy.

\subsection{Energy upper bound}

We prove an upper bound to $E(N)$ by computing the energy of a suitable trial function. The main idea is to modify the trial state $\psi_\mathrm{gauss}$ from \eqref{eq:psi_gauss}, which has no excitation content (it is the vacuum for the excitation space), by inserting a suitable $\Phi\in\ell^{2}(\mathcal{F}_\perp)$ that captures the minimal excitation energy $E^\mathrm{Bog}$. The 2-mode part of $\psi_\mathrm{gauss}$, that is, the relative weights of the occupation of $u_1$ with respect to that of $u_2$, should not be changed, since we showed in Proposition \ref{prop:H_2-mode} that it gives an upper bound for $H_{2-\mathrm{mode}}$ that matches the lower bound.

The excitation content of the trial state has to be fixed in order to capture the minimal energy of $\mathbb{H}_\mathrm{right}^{( M)}$ and $\mathbb{H}_\mathrm{left}^{( M)}$ from \eqref{eq:H_bog_cutoff}. It turns out that it is in fact enough to consider the optimal trial state for the cutoff-less versions of $\mathbb{H}_\mathrm{right}^{(M)}$ and $\mathbb{H}_\mathrm{left}^{(M)}$. Moreover, a further simplification is possible, namely to formally replace the $\Theta$ operator by the identity. This will have a negligible effect on the energy of the trial state. Overall, we are looking for an excitation vector that optimizes the operator
\begin{equation*}
\begin{split}
\mathbb{H}_\mathrm{right}^{\Theta=\mathbbm{1}}:=\;&\sum_{\alpha,\beta\ge1}\left\langle u_{r,\alpha},\Big(h_\mathrm{MF}-\mu_++{\lambda}K_{11}\Big) u_{r,\beta}\right\rangle  a^\dagger _{r,\alpha} a_{r,\beta}\\
&+\frac{\lambda}{2}\sum_{\alpha,\beta\ge1}\left\langle u_{r,\alpha}, K_{11} u_{r,\beta}\right\rangle \left( a^\dagger _{r,\alpha} a^\dagger _{r,\beta}+ a_{r,\alpha} a_{r,\beta}\right)
\end{split}
\end{equation*}
and the corresponding (commuting) operator
\begin{equation*}
\begin{split}
\mathbb{H}_\mathrm{left}^{\Theta=\mathbbm{1}}:=\;&\sum_{\alpha,\beta\ge1}\left\langle u_{\ell,\alpha},\Big(h_\mathrm{MF}-\mu_++{\lambda}K_{22}\Big) u_{\ell,\beta}\right\rangle  a^\dagger_{\ell,\alpha} a_{\ell,\beta}\\
&+\frac{\lambda}{2}\sum_{\alpha,\beta\ge1}\left\langle u_{\ell,\alpha}, K_{22} u_{\ell,\beta}\right\rangle \left( a^\dagger _{\ell,\alpha} a^\dagger _{\ell,\beta}+ a_{\ell,\alpha} a_{\ell,\beta}\right).
\end{split}
\end{equation*}

It is well-known that Hamiltonians of this type are explicitly diagonalized by a unitary operator that implements a Bogoliubov transformation, and that the eigenvector corresponding to the lowest eigenvalue is obtained by acting on the vacuum with such a diagonalizing unitary operator. This means that there exist two commuting (since they are constructed using orthogonal modes) unitary operators $\mathbb{U}_\mathrm{right}$ and $\mathbb{U}_\mathrm{left}$ such that
\begin{equation*}
\begin{split}
\mathbb{H}_\mathrm{right}^{\Theta=\mathbbm{1}} \mathbb{U}_\mathrm{right}\Omega=\;&E^\mathrm{Bog}_{\mathrm{right}} \mathbb{U}_\mathrm{right}\Omega,\\
\mathbb{H}_\mathrm{left}^{\Theta=\mathbbm{1}} \mathbb{U}_\mathrm{left}\Omega=\;&E^\mathrm{Bog}_{\mathrm{left}} \mathbb{U}_\mathrm{left}\Omega
\end{split}
\end{equation*}
for some eigenvalues $E^\mathrm{Bog}_{\mathrm{right}}$ and $E^\mathrm{Bog}_{\mathrm{left}}$. Moreover, it turns out that (see \cite[Section 8.1]{OlgRouSpe-21})
\begin{equation*}
E^\mathrm{Bog}_{\mathrm{right}}+E^\mathrm{Bog}_{\mathrm{left}}=E^\mathrm{Bog}
\end{equation*}
where $E^\mathrm{Bog}$ is the quantity defined in \eqref{eq:E^bog}.

We are now ready to define our trial state. Let us define the coefficients
\begin{equation} \label{eq:c_d_again}
c_{d,s}=
\begin{cases}
\frac{1}{Z_N}e^{-d^2/4\sigma_N^2} & \text{if $N-s+d$ is even and $|d| \le \sigma_N^2$} \\
0                           & \text{otherwise,}
\end{cases}  
\end{equation}
where $Z_N$ is a normalization factor ensuring that $\sum_{|d|\le \sigma_N^2} c^2_{d,s}=1$, and the variance $\sigma_N^2$ is given by
\begin{equation}\label{eq:modif sigma}
\sigma_N ^2 = \begin{cases}
\sqrt{\mu_- - \mu_+} N \quad \mbox{ if } \delta < 1 \mbox{ in the assumption }T\sim N^{-\delta} \\
N^{1/2} \quad \mbox{ otherwise.} 
\end{cases}
\end{equation}
Let us also define the excitation vector
\begin{equation} \label{eq:phi_trial}
\Phi_{\mathrm{trial},s}:= \frac{\big(\mathbb{U}_\mathrm{left} \mathbb{U}_\mathrm{right} \Omega\big)_{s}}{\sqrt{\sum_{s=0}^N\left\|\big(\mathbb{U}_\mathrm{left} \mathbb{U}_\mathrm{right} \Omega\big)_{s}\right\|^2}} \in\mathcal{F}_\perp.
\end{equation}
We then set
\begin{equation}
\psi_\mathrm{trial}:=\sum_{s=0}^N \sum_{|d|\le \sigma_N^2} c_{d,s}u_1^{\otimes(N-s+d)/2}\otimes_{\mathrm{sym}} u_2^{\otimes (N-s-d)/2}\otimes_{\mathrm{sym}}\Phi_{\mathrm{trial},s}.
\end{equation}

\begin{proposition}[\textbf{Energy upper bound}]
	Assume $T\sim N^{-\delta}$ for some $\delta>0$. Then
	\begin{equation*}
	\limsup_{N\to\infty}\big(\langle H_N\rangle_{\psi_\mathrm{trial}}-E_{2-\mathrm{mode}}-E^\mathrm{Bog}\big)\le 0.
	\end{equation*}
\end{proposition}

We refer the reader to \cite[Section 8.1]{OlgRouSpe-21} for the proof. The overall strategy goes through an application of the results of Section \ref{sect:2_mode}, \ref{sect:excitations}, and \ref{sect:diagonalization}, after noticing that $\psi_\mathrm{trial}$ satisfies all the assumptions of those intermediate results. This is quite similar to what we do for the lower bound (except with $\psi_\mathrm{gs}$ replacing $\psi_\mathrm{trial}$), and we will thus discuss it in greater detail in that case.

\subsection{Energy lower bound}

To obtain an energy lower bound we apply the results of Section \ref{sect:2_mode}, \ref{sect:excitations}, and \ref{sect:diagonalization} after computing the expectation on the ground state $\psi_\mathrm{gs}$. This will give a lower bound on $E(N) = \langle H_N\rangle_{\psi_\mathrm{gs}}$.

First, we notice that due to the a priori bounds of Proposition \ref{prop:H_2-mode}, $\psi_\mathrm{gs}$ satisfies the assumptions of Proposition \ref{prop:bogoliubov}. This implies
\begin{equation}
\begin{split}
\langle H_N\rangle_{\psi_\mathrm{gs}} \ge\;&  \langle H_{2-\mathrm{mode}}\rangle_{\psi_\mathrm{gs}}+\mu_+\langle \mathcal{N}_\perp \rangle_{\psi_\mathrm{gs}} +\langle \mathbb{H}\rangle_{ U_N\psi_\mathrm{gs}}\\
&+\frac{\lambda}{\sqrt{2(N-1)}}\sum_{m\ge3}\Big\langle U_N \psi_\mathrm{gs}, \Big(\big(w_{+1-m}\,\Theta+w_{+2-m}\, \Theta^{-1}\big) a_m\mathfrak{D}+\mathrm{h.c.}\Big)U_N\psi_\mathrm{gs}\big\rangle\\
&-o_N(1).
\end{split}
\end{equation}
We now apply the results of Proposition \ref{prop:reduction} to replace, in the right hand side, $\mathbb{H}$ and the linear terms by operators coupling only right- or left-well modes with themselves. Again, the a priori bounds on $\psi_\mathrm{gs}$ ensure that the assumptions of Proposition \ref{prop:reduction} are satisfied. We then have, for any energy cutoff $\Lambda$ large enough,
\begin{equation*}
\begin{split}
\langle H_N\rangle_{\psi_\mathrm{gs}}\ge \;& \langle H_{2-\mathrm{mode}}\rangle_{\psi_\mathrm{gs}}+\mu_+\langle \mathcal{N}_\perp \rangle_{\psi_\mathrm{gs}}+ \Big\langle \mathbb{H}_\mathrm{right}^{(M_\Lambda)}+\mathbb{H}_\mathrm{left}^{(M_\Lambda)}\Big\rangle_{ U_N\psi_\mathrm{gs}}\\
&+\frac{\lambda}{\sqrt{2(N-1)}}\sum_{1\le \alpha \le M_\Lambda} \big\langle u_1, w*(u_+u_-) u_{r,\alpha}\big\rangle\,\big\langle b_{r,\alpha}\mathfrak{D}+\mathrm{h.c.}\big\rangle_{U_N \psi_\mathrm{gs}}\\
&+\frac{\lambda}{\sqrt{2(N-1)}}\sum_{1\le \alpha \le M_\Lambda} \big\langle u_2, w*(u_+u_-) u_{\ell,\alpha}\big\rangle\,\big\langle c_{\ell,\alpha}\mathfrak{D}+\mathrm{h.c.}\big\rangle_{U_N \psi_\mathrm{gs}}\\
&-C_\Lambda o_N(1)-\frac{C}{(\mu_{2M_\Lambda+2}-\mu_+)}^{1/2}.
\end{split}
\end{equation*}
The next step is to use the lower bound from Proposition \ref{prop:lower_shifted} on the Bogoliubov Hamiltonians together with the linear terms. This gives
\begin{equation*}
\begin{split}
\langle H_N\rangle_{\psi_\mathrm{gs}}\ge \;& \langle H_{2-\mathrm{mode}}\rangle_{\psi_\mathrm{gs}}+\mu_+\langle \mathcal{N}_\perp \rangle_{\psi_\mathrm{gs}}+E^\mathrm{Bog}\\
&-\frac{\lambda^2}{2(N-1)} \Big(\big\langle u_1,K_{11}W_{r,\le M_\Lambda} K_{11} u_1\big\rangle+\big\langle u_2,K_{22}W_{\ell,\le M_\Lambda} K_{22} u_2\big\rangle\Big) \langle (\mathcal{N}_1-\mathcal{N}_2)^2\rangle_{\psi_\mathrm{gs}}\\
&-C_\Lambda o_N(1)-\frac{C}{(\mu_{2M_\Lambda+2}-\mu_+)}^{1/2}.
\end{split}
\end{equation*}
We still have to extract the leading order energy terms from $H_{2-\mathrm{mode}}$. This is achieved using the lower bound \eqref{eq:H_2-mode_lower} as well as \eqref{eq:E_2_mode_bound}. We combine this with the lower bound \eqref{eq:lower_bound_coefficient} on the coefficient of the variance operator which holds if the coupling constant $\lambda$ is small enough. We get
\begin{equation*}
\begin{split}
\langle H_N\rangle_{\psi_\mathrm{gs}}\ge \;& E_{2-\mathrm{mode}}+\frac{c \lambda }{N}\big\langle(\mathcal{N}_1-\mathcal{N}_2)^2\big\rangle_{\psi_\mathrm{gs}}+E^\mathrm{Bog}\\
&-C_\Lambda o_N(1)-\frac{C}{(\mu_{2M_\Lambda+2}-\mu_+)}^{1/2}.
\end{split}
\end{equation*}
We have thus proven the following result.

\begin{proposition}[\textbf{Energy lower bound}]
	Assume $\lim_{N\to\infty} T=0$. For any energy cutoff $\Lambda$ large enough, let $M_\Lambda$ be the largest integer such that $\mu_{2M_\Lambda+2}\le \Lambda$, where $\{\mu_m\}$ are the eigenvalues of $h_\mathrm{MF}$ in increasing order. Then, assuming $0<\lambda<\lambda_0$ for some $\lambda_0>0$ small enough, we have
	\begin{equation} \label{eq:lower_bound}
	\begin{split}
	\langle H_N\rangle_{\psi_\mathrm{gs}}\ge \;& E_{2-\mathrm{mode}}+\frac{c \lambda }{N}\big\langle(\mathcal{N}_1-\mathcal{N}_2)^2\big\rangle_{\psi_\mathrm{gs}}+E^\mathrm{Bog}\\
	&-C_\Lambda o_N(1)-\frac{C}{(\mu_{2M_\Lambda+2}-\mu_+)}^{1/2},
	\end{split}
	\end{equation}
	for a small enough but strictly positive constant $c$.
\end{proposition}

\subsection{Proof of Theorem \ref{thm:main}}

Passing to the limit in \eqref{eq:lower_bound} we have
\begin{equation*}
\liminf_{N\to\infty} \left( \langle H_N \rangle_{\psi_\mathrm{gs}}- E_{2-\mathrm{mode}}-E^\mathrm{Bog} \right) \ge \limsup_{N\to\infty}  \left(\frac{c \lambda }{N}\big\langle(\mathcal{N}_1-\mathcal{N}_2)^2\big\rangle_{\psi_\mathrm{gs}}-\frac{C}{(\mu_{2M_\Lambda+2}-\mu_+)}^{1/2} \right).
\end{equation*}
On the other hand, the energy upper bound implies
\begin{equation*}
\limsup_{N\to\infty}\left( \langle H_N \rangle_{\psi_\mathrm{gs}}- E_{2-\mathrm{mode}}-E^\mathrm{Bog} \right) \le0.
\end{equation*}
Now, it can be seen as an extension of our results in \cite[Theorem 2.2]{OlgRouSpe-21} that the quantity $\mu_{2M_\Lambda+2}$ has a limit as $N\to\infty$. Such a limit coincides with the $(M_\Lambda+1)$-th eigenvalue of a one-well mean-field Hamiltonian, which is a quantity that diverges as $\Lambda\to\infty$. This implies
\begin{equation*}
\limsup_{N\to\infty} \frac{c\lambda}{N}\big\langle(\mathcal{N}_1-\mathcal{N}_2)^2\big\rangle_{\psi_\mathrm{gs}} \le o_\Lambda (1),
\end{equation*}
and passing now to the $\Lambda\to\infty$ limit (recalling that $(\mathcal{N}_1-\mathcal{N}_2)^2\ge0$) proves \eqref{eq:main_variance}. The energy expansion \eqref{eq:main_energy} then follows once again from the upper and lower bounds, using now the fact that the variance term converges to zero.

\subsection*{Acknowledgments} The author gratefully acknowledges the support from the European Research Council through the ERC-StG CORFRONMAT and the ERC-AdG CLaQS. Fruitful discussions with S.~Cenatiempo and S.~Rademacher, as well as with the coauthors of~\cite{OlgRouSpe-21} N.~Rougerie and D.~Spehner are also warmly acknowledged.

\end{document}